\documentclass[conference]{IEEEtran}
\IEEEoverridecommandlockouts

\usepackage{cite}
\usepackage{amsmath,amssymb,amsfonts}
\usepackage{graphicx}
\usepackage{booktabs}

\usepackage{algorithm}
\usepackage{algpseudocode}

\usepackage{textcomp}
\usepackage{xcolor}
\usepackage[hidelinks]{hyperref}

\def\BibTeX{{\rm B\kern-.05em{\sc i\kern-.025em b}\kern-.08em
    T\kern-.1667em\lower.7ex\hbox{E}\kern-.125emX}}
\begin{document}

\title{Optimizing LLM Code Suggestions: Feedback-Driven Timing with Lightweight State Bounds}

\author{\IEEEauthorblockN{Mohammad Nour Al Awad}
	\IEEEauthorblockA{
		\textit{ITMO University}\\
		Saint Petersburg, Russia \\
		MohammadNourAlAwad@itmo.ru}
	\and
	\IEEEauthorblockN{Sergey Ivanov}
	\IEEEauthorblockA{
		\textit{ITMO University}\\
		Saint Petersburg, Russia \\
		svivanov@itmo.ru}
	\and
	\IEEEauthorblockN{Olga Tikhonova}
	\IEEEauthorblockA{
		\textit{ITMO University}\\
		Saint Petersburg, Russia \\
		tikhonova\_ob@itmo.ru}
}

\maketitle
\begingroup
\renewcommand\thefootnote{}\footnote{
	© 2025 IEEE. Personal use of this material is permitted.
	Permission from IEEE must be obtained for all other uses.
}\addtocounter{footnote}{-1}
\endgroup

\begin{abstract}
	Large Language Models (LLMs) have transformed code auto-completion by generating context-aware suggestions. Yet, deciding \emph{when} to present these suggestions remains underexplored, often leading to interruptions or wasted inference calls. We propose an adaptive timing mechanism that dynamically adjusts the delay before offering a suggestion based on real-time developer feedback. Our suggested method combines a logistic transform of recent acceptance rates with a bounded delay range, anchored by a high-level binary prediction of the developer’s cognitive state. In a two-month deployment with professional developers, our system improved suggestion acceptance from 4.9\% with no delay to 15.4\% with static delays, and to 18.6\% with adaptive timing—while reducing blind rejections (rejections without being read) from 8.3\% to 0.36\%. Together, these improvements increase acceptance and substantially reduce wasted inference calls by 75\%, making LLM-based code assistants more efficient and cost-effective in practice.		
\end{abstract}

\begin{IEEEkeywords}
	Behavioral Modeling, Adaptive Systems, Interaction timing, Human-Computer Interaction, Code Completion
\end{IEEEkeywords}

\section{Introduction}
Modern software development increasingly relies on AI-powered code assistants—most prominently LLM-based tools such as GitHub Copilot—which leverage massive pre-trained models to suggest context-aware completions and entire code snippets \cite{brown2020language,chen2021evaluating}. These systems aim to boost productivity by reducing boilerplate and aiding API recall. Subsequent work on specialized code models (e.g., CodeBERT and CodeT5) has further improved completion accuracy and relevance \cite{feng2020codebert,wang2021codet5, wang2024systematic}.

Despite advances in \emph{what} content to generate, the \emph{timing} of suggestion delivery remains an underexplored yet critical factor. Immediate, keystroke-triggered suggestions can interrupt a developer’s flow or be ignored outright, while overly delayed prompts risk becoming irrelevant once the context has changed. Mistimed suggestions not only degrade user experience but also waste expensive LLM inference calls.

In this paper, we focus squarely on optimizing suggestion timing. We introduce a lightweight, feedback-driven algorithm that modulates delay based on real-time acceptance rates, bounded by a minimal binary prediction of the developer’s cognitive state (implementing vs.\ debugging). Our contributions are:

\begin{itemize}
	\item A formal, bounded delay adaptation mechanism using a logistic mapping of acceptance history.
	\item Integration with a cognitive state classifier to anchor timing ranges.
	\item A real-world deployment showing meaningful 3x increases in suggestion acceptance and substantial reductions in wasteful completions.
\end{itemize}

The paper is organized as follows. Section 2 surveys related work on code suggestion models and adaptive human–computer interaction. Section 3 details our adaptive delay method, including the acceptance-rate transform and state anchoring. Section 4 describes the system architecture and experimental design, and Section 5 presents the deployment results with a discussion of efficiency and user experience. Finally, Section 6 concludes and outlines directions for future research.

\section{Related Work}

\subsection{LLM Code Completion and Programming Assistance}

Recent advances in large language models (LLMs) have significantly improved the accuracy and fluency of code completion systems. Early breakthroughs such as GPT-3~\cite{brown2020language} demonstrated the capacity of general-purpose transformers to perform few-shot learning across diverse code-related tasks. This was followed by specialized models like Codex~\cite{chen2021evaluating}, CodeBERT~\cite{feng2020codebert}, and CodeT5~\cite{wang2021codet5}, which introduced inductive biases for programming syntax and identifier structure, yielding improved performance on both generation and understanding benchmarks.

More recently, open-source efforts such as StarCoder~\cite{li2023starcoder}, Code LLaMA~\cite{wang2023codellama}, and CodeChain~\cite{chen2023codechain} have scaled token context windows, improved multilingual coverage, explored chain-of-thought prompting for complex tasks and domain‑adaptive techniques~\cite{tang2023domain}. Recent robustness work also studies adversarial prompt design for LLM completer \cite{li2024advpro}. While foundational for commercial assistants, these models focus on suggestion content rather than timing or interaction dynamics. Meanwhile, holistic benchmarks such as ComplexCodeEval~\cite{feng2024complexcodeeval} highlight the need for richer, context‑aware assessment of code‑oriented LLMs.

Despite these advances, most systems remain limited to static invocation triggers (e.g., after every keystroke or upon user request), overlooking the cognitive and behavioral variability of the developer. Our work focuses on this under explored temporal axis of adaptivity.

\subsection{Suggestion Timing and Cognitive Load in Code Assistance}

The question of \emph{when} to offer a code suggestion has only recently received attention in the literature. Empirical studies have shown that poorly timed suggestions can introduce cognitive overhead, disrupt developer flow, or even reduce trust in the assistant~\cite{sida2023copilot,barke2022grounded,ziegler2022productivity}. For example,~\cite{barke2022grounded} reported that aggressive suggestion timing often leads developers to immediately dismiss completions without reading them—an effect our system explicitly targets via blind rejection reduction. Recent simulation studies such as RepoSim corroborate this observation by replaying realistic typing traces \cite{peng2024reposim}.

Cognitive science research has long emphasized the cost of interruptions in high-load tasks, yet most current assistants apply fixed or heuristically determined delays. Recent proposals such as~\cite{mozannar2023when} and~\cite{pu2025assistance} explore proactive suggestion timing, proposing conceptual frameworks for timing-aware interactions, but do not implement bounded, feedback-driven controllers in real-world systems.

The closest implementation-level work is~\cite{demoor2024transformer}, which proposes a transformer-based classifier to predict opportune moments for suggestion invocation. While promising, their system does not integrate developer feedback into a control loop or deploy in production. We address this gap by deploying a bounded adaptation policy that adjusts timing based on developer state and real-time feedback.

\subsection{Adaptive Interfaces and Human-in-the-Loop Feedback Loops}

Traditional systems adjusted interface behaviors using explicit or implicit user state, ranging from idle detection to eye-tracking. In the LLM era, adaptivity has largely focused on content generation—e.g., via reinforcement learning with human feedback (RLHF)~\cite{ouyang2022training,stiennon2020learning}, prompt evolution~\cite{chen2023promptbreeder}, or dynamic sampling policies~\cite{su2024dynamic}.

Most focus on optimizing content generation, with little attention to the timing of suggestion delivery. Adaptive timing remains a relatively untouched axis despite its central role in user satisfaction and efficiency. Our approach complements these content-centric strategies by modeling and controlling the delivery timing as a function of user response, offering a lightweight yet powerful feedback loop without requiring fine-grained RL policies.

\subsection{System Efficiency and Real-World Deployments}

Few prior studies have examined LLM code assistants in long-term production settings with infrastructure cost and user experience metrics in mind. Most evaluation studies, such as~\cite{sida2023copilot,ziegler2022productivity}, rely on surveys, simulated tasks, or telemetry snapshots rather than real-world deployments with controlled variable manipulation. As a result, the interaction between suggestion frequency, user acceptance, and backend resource usage remains poorly quantified.

Our work directly contributes to this emerging need for rigorous evaluation of production-side implications. By framing timing adaptation as a bounded control policy, we guarantee predictable worst-case inference rates—crucial for GPU provisioning in real systems. Moreover, our method directly ties delay changes to acceptance feedback, closing the loop between user behavior and system cost.

While some works (e.g.,~\cite{pu2025assistance}) touch on the productivity tradeoffs of suggestion timing, they stop short of implementing adaptive systems or quantifying cost-efficiency improvements. Our results demonstrate that suggestion acceptance can be tripled while reducing backend load by 75\%—an unprecedented efficiency gain in this domain.

\smallskip
To our knowledge, this is the first work to implement and rigorously evaluate a production-grade, behaviorally adaptive timing controller for LLM code assistants. We integrate lightweight telemetry, cognitive state prediction, and a mathematically bounded feedback loop to regulate suggestion delays in real time. Our deployment shows that timing—when treated as a controllable design parameter—can substantially improve both usability and infrastructure efficiency. This work opens a new axis of adaptivity that complements content optimization and enables safer, more cost-effective integration of LLMs into everyday developer workflows.

\section{Adaptive Delay Method}\label{sec:delay_method}

Our adaptive suggestion timing mechanism is driven primarily by recent suggestion acceptance feedback. To keep the delay within sensible limits, we anchor it using a binary prediction of the developer’s high-level cognitive state.

\subsection{High-level State Anchoring}\label{sec:state}

Timing code suggestions well requires not only short-term behavioral feedback (like acceptance rate) but also a stable signal about the developer’s current cognitive context. Acceptance-based adaptation alone can misfire if it ignores the shift in developer tasks with very different interruption tolerances. Our system therefore anchors its adaptive delay using a minimal but effective binary developer state: \emph{implementing} vs.\ \emph{debugging}.

This high-level state classifier plays an important role: it sets strict upper and lower bounds for the delay range (Table~\ref{tab:delay_ranges}). While the acceptance rate transform dynamically tightens or relaxes the delay, the state ensures that suggestions cannot become too aggressive when the developer is debugging — or unnecessarily sluggish when actively implementing. Without this guardrail, the system could easily produce mistimed completions that disrupt flow or miss helpful moments.

\subsubsection{Design Motivation and Practical Tradeoffs}

Early pilots explored more granular state models. Developers initially labeled their own sessions with up to seven detailed states, including testing, reading documentation, and refactoring. However, these fine-grained labels quickly proved unreliable and highly imbalanced: developers found real-time labeling tedious, leading to noisy or inconsistent categories.

Grouping states into just two broad modes — \emph{implementing} and \emph{debugging} — gave a far more robust signal that still captured the dominant effect on suggestion timing. Developers tolerate short, frequent completions while writing or expanding code, but are more sensitive to interruptions when inspecting logs, stepping through errors, or searching for root causes.

\subsubsection{Lightweight LLM-Based Classification}

To infer the developer state with minimal overhead, we first trained gradient-boosted tree models (e.g., XGBoost) using our collected IDE telemetry metrics (typing cadence, navigation patterns, command usage, idle time). These models reached about 75\% cross-validated accuracy on the developer-labeled ground truth (binary classification after grouping the original labels). However, this accuracy was limited by the inconsistency of the original developer-provided labels. To improve reliability, we manually annotated a smaller, high-quality subset of 100 real intervals from the data. Using this refined data, we then tested a compact prompt-based LLM classifier that summarizes one minute of activity together with the \textbf{code changes (diff)} as a short natural language prompt. The LLM approach exceeded the tree model’s accuracy (over 84\% agreement with the manually labeled subset) while requiring no model retraining or additional local compute. This highlights the LLM’s capacity to infer cognitive state by combining behavioral metrics with an understanding of the code diff itself.

This classifier is stateless and unobtrusive: it runs once per minute, sends only minimal summary signals. This makes it practical for real-world IDE deployments without major performance tradeoffs.

\subsubsection{Impact on Timing Stability}

The high-level state does not control when to show a suggestion directly — instead, it sets the safe range within which the acceptance feedback loop operates. For example, when the system detects that the developer is debugging, it shifts the minimum suggestion delay to 1.0–1.6 seconds, ensuring that no premature completions interrupt focused problem-solving. During active implementation, the range shifts to 0.8–1.4 seconds, allowing more responsive suggestions when typing pauses naturally occur.

Together with the logistic transform and smoothing range, this binary signal guarantees that the adaptive delay remains stable, predictable, and aligned with the developer’s moment-to-moment mental load — while avoiding unnecessary complexity or intrusive behavioral tracking.

\subsection{Acceptance-Rate Based Delay Adaptation}\label{sec:method}

Once the developer’s high-level state defines a safe timing range, the system adapts the precise delay within that range by monitoring real-time suggestion acceptance. This second layer turns recent developer behavior into a direct feedback loop that tightens or relaxes the delay to match how well suggestions are received.

Below we present the final production formulation deployed in our system. This algorithm (i) respects the state-specific bounds from Table~\ref{tab:delay_ranges}, (ii) uses a symmetric logistic transform to smoothly map acceptance rates to delay adjustments, and (iii) includes a per-update delay smoothing to avoid abrupt swings.

\smallskip
\subsubsection{\textbf{Base Delay Ranges}}

To determine appropriate base delays for each developer state, we analyzed real-world typing behavior during the deployment. For each one-minute interval of active development, we filtered segments where typing speed ranged between 0 and 5 characters per second—excluding non-typing actions like copying or suggestion acceptance. From these, we computed inter-keystroke intervals as the inverse of the average typing speed.

Analysis shows that during debugging, inter-keystroke intervals were notably longer than during implementation, reflecting greater cognitive effort. We therefore selected the 97th percentile of the observed typing intervals to define the safe delay estimation for each state, yielding 1.068 seconds for implementing and 1.293 seconds for debugging. For consistency and ease of tuning, we rounded these to define the base delays: \(D_{\mathrm{base}} = 1.1\,\text{s}\) for implementing and \(1.3\,\text{s}\) for debugging.

To construct bounded timing ranges, we applied a symmetrical ±0.3\,s margin around each base delay, resulting in the ranges shown in Table~\ref{tab:delay_ranges}. These were designed to exceed typical pause durations, minimizing the likelihood of \emph{blind rejections}—cases where suggestions are dismissed immediately because they arrive just as the developer resumes typing.

\begin{table}[h]
	\caption{Delay ranges by predicted developer state.}
	\label{tab:delay_ranges}
	\centering
	\begin{tabular}{lccc}
		\toprule
		\textbf{State} & \(D_{\min}\) (s) & \(D_{\max}\) (s) & \(D_{\mathrm{base}}\) (s) \\
		\midrule
		implementing & 0.80 & 1.40 & 1.10 \\
		debugging    & 1.00 & 1.60 & 1.30 \\
		\bottomrule
	\end{tabular}
\end{table}

We also compute a dimensionless gain factor
\begin{equation*}
	K = \frac{D_{\max} - D_{\min}}{D_{\max} + D_{\min}}
\end{equation*}
which is approximately 0.273 for \emph{implementing} and 0.231 for \emph{debugging}.  \(K\) determines how aggressively we stretch or shrink \(D_{\mathrm{base}}\).

\smallskip
\subsubsection{\textbf{Acceptance-Rate Transform}}

From the last minute of interaction, we compute the acceptance rate
\begin{equation*}
	\begin{split}
		A &= \frac{n_{\mathrm{accepted}}}{n_{\mathrm{accepted}} + n_{\mathrm{rejected}}},\\
		A &= 0 \quad\text{if }n_{\mathrm{accepted}} + n_{\mathrm{rejected}} = 0
	\end{split}
\end{equation*}

To map this bounded signal into a symmetric shape that is steep near a reference value and saturates at the extremes, we use a scaled logistic function because it provides smooth, bounded adjustments that are highly sensitive around the inflection point while naturally saturating at the extremes, preventing unstable delay swings.

\begin{equation*}
	\begin{split}
		L(x) &= \frac{2}{1 + e^{-x}} - 1,\quad L(x)\in(-1,1),\\
		S_{\mathrm{raw}} &= L\bigl(\gamma (A - A_0)\bigr)
	\end{split}
\end{equation*}

\noindent Here, $\gamma=10$ and $A_0=0.15$, to make the logistic transform responsive in the mid-range without saturating at realistic acceptance rates ($A\in[0.05,0.40]$). The inflection $A_0{=}0.15$ reflects the point where developers begin to perceive suggestions as distracting, and we set $\gamma=10$ to tune sensitivity of the changes in acceptance rate.

Because \(L\) is not perfectly anti-symmetric over \([-\gamma A_0, \gamma(1 - A_0)]\), we normalize to the full \([-1,1]\) range:
\begin{equation*}
	S_0 = L\bigl(-\gamma A_0\bigr), \quad S_1 = L\bigl(\gamma (1 - A_0)\bigr)
\end{equation*}

\begin{equation*}\label{eq:Snorm}
	S_{\mathrm{norm}}
	= \frac{2\bigl(S_{\mathrm{raw}} - S_0\bigr)}{S_1 - S_0} - 1
\end{equation*}

Thus, \(S_{\mathrm{norm}} = -1\) when \(A=0\) and \(+1\) when \(A=1\).

\smallskip
\subsubsection{\textbf{Delay Adjustment Formula}}

The multiplicative factor applied to the base delay is
\begin{equation*}\label{eq:F}
	F = 1 - K\,S_{\mathrm{norm}}
\end{equation*}

so that the predicted delay is
\begin{equation*}
	\begin{split}
		D_{\mathrm{pred}} &= D_{\mathrm{base}} \times F,\\
		D_{\mathrm{pred}} &\in \bigl[(1 - K)\,D_{\mathrm{base}},\,(1 + K)\,D_{\mathrm{base}}\bigr].
	\end{split}
\end{equation*}

For example, in an \emph{implementing} interval with \(D_{\mathrm{base}} = 1.10\,\mathrm{s}\) and \(K = 0.273\):
\begin{itemize}
	\item \(A=1\):
	\(F \approx 0.727 \;\Rightarrow\; D_{\mathrm{pred}}\approx 0.80\,\mathrm{s}\).
	\item \(A=0\):
	\(F \approx 1.273 \;\Rightarrow\; D_{\mathrm{pred}}\approx 1.40\,\mathrm{s}\).
\end{itemize}

\begin{figure}[h]
	\centering
	\includegraphics[width=0.9\linewidth]{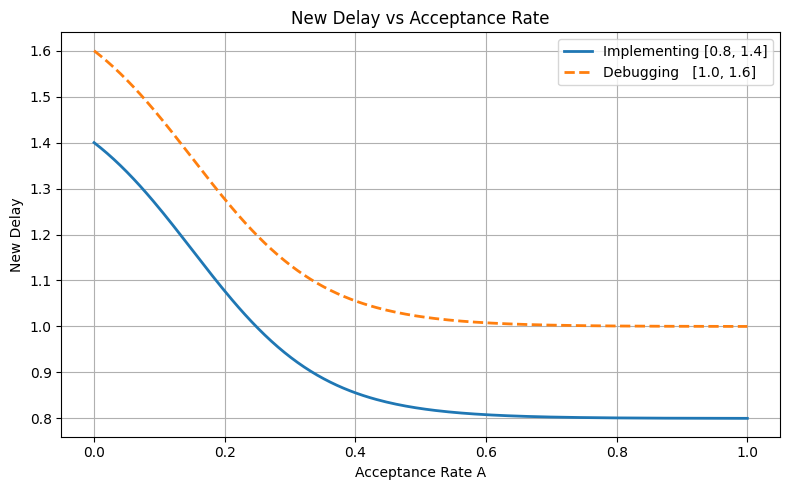}
	\caption{Effect of the acceptance rate $A$ on the computed delay $D_{\mathrm{pred}}$ for both predicted states. The logistic transform smoothly shifts the delay within its bounded range, tightening the timing for higher acceptance and relaxing it for lower acceptance.}
	\label{fig:delay_vs_acceptance}
\end{figure}

Thus, the delay remains strictly bounded within empirically defined safe limits.
Figure~\ref{fig:delay_vs_acceptance} shows how the final delay varies with the acceptance rate for both predicted states. This illustrates how our logistic mapping smoothly tightens the timing when recent suggestions are well-received, while ensuring the delay stays within safe bounds.

\smallskip
\subsubsection{\textbf{Per-Update Smoothing}}

To eliminate audible “pops” in the UI when the delay would otherwise change by hundreds of milliseconds between successive minutes, we apply a first-order delay range:
\begin{equation*}\label{eq:delta}
	\Delta = \mathrm{clip}\bigl(D_{\mathrm{pred}} - D_{\mathrm{old}},\,-0.10,\,+0.10\bigr)
\end{equation*}

\begin{equation*}\label{eq:Dnew}
	D_{\mathrm{new}} = D_{\mathrm{old}} + \Delta
\end{equation*}

With the \(\pm0.10\,\mathrm{s}\) ceiling, it takes at least 5~minutes for a change to propagate from one extreme of the band to the other, providing a smooth, user-imperceptible transition.

\begin{algorithm}[t]
	\caption{Adaptive Delay Controller}
	\label{alg:adaptive_delay}
	\begin{algorithmic}[1]
		\Require \\
		
		Developer state $s\!\in\!\{\textsc{Implementing},\textsc{Debugging}\}$\\
		minute–level acceptance counts $(n_{\mathrm{acc}}, n_{\mathrm{rej}})$\\
		previous delay $D_{\text{old}}$
		\Ensure  New delay $D_{\text{new}}$
		\State $(D_{\min},D_{\max}) \gets$ \textproc{StateRange}$(s)$      \Comment{Table \ref{tab:delay_ranges}}
		\State $D_{\text{base}} \gets (D_{\min}+D_{\max})/2$
		\State $K \gets (D_{\max}-D_{\min})/(D_{\max}+D_{\min})$
		\State $A \gets
		\begin{cases}
			0,& n_{\mathrm{acc}}+n_{\mathrm{rej}}=0\\
			\frac{n_{\mathrm{acc}}}{n_{\mathrm{acc}}+n_{\mathrm{rej}}},&\text{otherwise}
		\end{cases}$                                             
		\State $S_{\text{raw}} \gets 2/(1+\exp(-\gamma(A-A_0)))-1$           \Comment{$\gamma{=}10$, $A_0{=}0.15$}
		\State $S_{\text{norm}} \gets$ \textproc{NormalizeTo$\pm1$}$(S_{\text{raw}})$
		\State $F \gets 1 - K\cdot S_{\text{norm}}$
		\State $D_{\text{pred}} \gets D_{\text{base}} \times F$
		\State $\Delta \gets \text{clip}(D_{\text{pred}}-D_{\text{old}},\,{-}0.10,\,{+}0.10)$
		\State $D_{\text{new}} \gets D_{\text{old}} + \Delta$
		\State \Return $D_{\text{new}}$
	\end{algorithmic}
\end{algorithm}

\smallskip
\subsubsection{\textbf{Practical Notes \& Impact}}

\begin{itemize}
	\item \textbf{Why the lower \(A_0\)?} Pilot logs showed that once acceptance dips below \(\approx10\%\), developers already perceive the assistant as distracting. Shifting the logistic inflection point downwards makes the model more conservative in that region.
	\item \textbf{Bounded output means bounded compute.} Because \(D_{\mathrm{pred}}\) never exits \([D_{\min},D_{\max}]\), the backend call rate is strictly capped, yielding deterministic worst-case load and additional GPU cost savings over aggressive suggestions.
	\item \textbf{Smoothing solves \emph{pogo sticking}.} Prior to the delay range, rapid swings caused some suggestions to arrive after the developer had resumed typing, artificially inflating rejection counts. The \(0.1\,\mathrm{s}\) cap eliminated that artifact.
\end{itemize}

Overall, this method provides a clean analytical guarantee on the delay range and a perceptually smooth evolution over time.

\subsection{System Architecture}

Our adaptive timing mechanism was deployed as part of a larger production-grade LLM-powered coding assistant. The system comprises a VS Code plugin. The plugin collects minimal behavioral metrics, schedules suggestions using the adaptive delay algorithm, and logs feedback on acceptances and rejections. The system handles real-time developer state classification and code suggestion generation, using Qwen models served via vLLM for low-latency inference. Although the full platform includes additional capabilities—such as personalized code indexing, retrieval-augmented generation (RAG), dynamic suggestion length adaptation, and advanced feedback logging—this paper focuses specifically on the timing adaptation component and its measured impact.

\section{Experimental Design and Deployment}\label{sec:experiment}

We conducted a real-world deployment study to rigorously evaluate the effect of our adaptive suggestion timing method in practice. This section details the participant profile, study phases, data collection, and evaluation plan.

\subsection{Participants}

Our adaptive suggestion system was deployed in a naturalistic academic research setting. Rather than recruiting a fixed participant group for a scripted study, we made the tool available to a pool of 25 researchers and developers via our internal communication channels. Lab members could freely install and use the plugin during their daily coding tasks. Usage was entirely voluntary and varied organically: some developers tried the system once or twice, while others used it regularly for extended periods.

Prior to the deployment, we surveyed 44 developers across multiple university research environments to identify those who regularly use Visual Studio Code as their primary IDE. From these, 25 were invited to participate and provided with instructions and technical support. Ultimately, 9 experienced developers actively used the assistant for a sufficient period, each contributing at least 50 recorded code suggestions. All active participants had at least 3 years of programming experience (with 78\% reporting over 5 years) and coded for more than 4 hours daily in most cases. Python was the dominant language among them, complemented by JavaScript, TypeScript, C++, and C. Notably, while all participants had prior experience with AI chat assistants (e.g., ChatGPT), most had not used AI-powered IDE plugins before, highlighting the novelty and practical relevance of the adaptive timing mechanism in real workflows.

All usage logs were anonymized and collected with informed consent in accordance with institutional policies.

\subsection{Experimental Phases}

We designed the deployment in three sequential phases to isolate the effect of timing:

\begin{itemize}
	\item 
	\textbf{Phase 1: No delay.} \\
	In the baseline phase, the system issued suggestions immediately after every keystroke. This represents the most aggressive (and common) mode in early commercial assistants, but typically leads to many premature completions.
	
	\item 
	\textbf{Phase 2: Static Delays.} \\
	In the second phase, we introduced fixed typing delays (0.6--1.4\,s depending on the day). This provided a high-level control to filter out obvious mistimed suggestions, serving as a realistic intermediate baseline.
	
	\item 
	\textbf{Phase 3: Adaptive Delays.} \\
	Finally, we deployed the full adaptive algorithm described in Section~\ref{sec:method}, dynamically adjusting delays using real-time acceptance feedback and cognitive state anchoring.
\end{itemize}

Each phase ran for at least 2 weeks, ensuring enough interaction data while minimizing participant fatigue or novelty effects. Developers continued normal daily tasks without artificial prompts.

\subsection{Data Collection}

The plugin continuously logged minimal behavioral telemetry needed to compute suggestion timing and acceptance metrics:

\begin{itemize}
	\item \textbf{Behavioral Signals.} 35 IDE activity metrics recorded per minute (typing speed, file edits, navigation, commands). These are also used by other modules in the larger system.
	\item \textbf{Suggestion Events.} For each suggestion: applied delay, acceptance or rejection, and decision time. Pilot studies showed that rejections within 0.3\,s consistently reflected blind dismissals, so we use this threshold throughout.
	\item \textbf{Developer State.} The high-level binary cognitive state (implementing/debugging) was inferred every minute using a small prompt to a lightweight LLM classifier.
\end{itemize}

All logs were anonymized and stored securely. While our full platform uses richer logs for advanced personalization (e.g., retrieval tuning, suggestion length), this study isolates timing to cleanly measure its solo effect.

\subsection{Evaluation Plan}

Our main evaluation metrics were:

\begin{itemize}
	\item \textbf{Suggestion Acceptance Rate:}
	\[
	A = \frac{n_{\mathrm{accepted}}}{n_{\mathrm{accepted}} + n_{\mathrm{rejected}}}.
	\]
	\item \textbf{Blind Rejection Ratio:}
	\[
	R_{\text{blind/reject}} = \frac{n_{\text{blind\_rejected}}}{n_{\mathrm{rejected}}}
	\]
	\[
	R_{\text{blind/suggest}} = \frac{n_{\text{blind\_rejected}}}{n_{\mathrm{total\_suggestions}}}
	\]
\end{itemize}

We primarily used descriptive statistics due to the small sample. While not claiming formal generalizability, this realistic deployment offers valuable in-situ evidence of how fine-grained timing interacts with developer behavior. We report clear relative trends and absolute gains across phases in Section~\ref{sec:results}.

\section{Results and Discussion}\label{sec:results}

This section presents the outcomes of our three-phase deployment: (1) no delay, (2) static delays, and (3) dynamically adaptive delays. We analyze how each design affected suggestion acceptance, blind rejection rates, and overall system efficiency. We also discuss how these findings relate to the broader capabilities of our production system and outline implications for multi-faceted adaptivity.

\subsection{Acceptance Rate Evolution}

\begin{table}[h]
	\caption{Suggestion-acceptance rates with standard errors (SE) and Wald 95 \% confidence intervals.}
	\label{tab:acceptance_ci}
	\centering
	\begin{tabular}{lccccc}
		\toprule
		\textbf{Phase} & $n$ & $k$ & \textbf{Rate (\%)} & \textbf{SE (\%)} & \textbf{95\% CI} \\
		\midrule
		No delay        & 5\,460 & 267 & 4.89 & 0.29 & [4.3, 5.5] \\
		Static Delay     & 1\,225 & 189 & 15.43 & 1.03 & [13.4, 17.5] \\
		Adaptive Delay   & 1\,032 & 192 & 18.60 & 1.21 & [16.2, 21.0] \\
		\bottomrule
	\end{tabular}
\end{table}

Table~\ref{tab:acceptance_ci} shows how acceptance rates evolved, where, $n$ is the number of suggestions and $k$ is the number accepted.
The sharp jump from 4.9\% to 15.4\% under static delays confirms that poorly timed suggestions are a major source of noise and user frustration. Adding our adaptive layer lifted acceptance further to 18.6\% — a modest absolute gain of 3.2\% but about 20\% relative to the static baseline. This illustrates that even after removing obvious mistiming, fine-grained behavioral adaptation yields measurable improvements.

The differing $n$ across phases reflects organic usage: Phase~1 was the default baseline and thus accumulated more suggestions; Phases~2–3 were time-boxed interventions with fewer total interactions.

\subsection{Blind Rejection Reductions}

We define blind rejections as suggestions dismissed almost instantly (within 0.3\,s). These often reflect suggestions that appear before the developer has paused long enough to read them. Table~\ref{tab:blind_rejections} summarizes the effect:

\begin{table}[h]
	\caption{Blind rejection rates by phase.}
	\label{tab:blind_rejections}
	\centering
	\begin{tabular}{lcc}
		\toprule
		\textbf{Phase} & \(\mathbf{R_{\text{blind/reject}}}\) & \(\mathbf{R_{\text{blind/suggest}}}\) \\
		\midrule
		No delay & 8.7\% & 8.3\% \\
		Static Delay & 1.3\% & 1.1\% \\
		Adaptive Delay & 0.36\% & 0.3\% \\
		\bottomrule
	\end{tabular}
\end{table}

The static delay alone reduced blind rejections by an order of magnitude. The adaptive mechanism sustained this low rejection rate while improving acceptance, showing that real-time feedback can safely tighten delays without regressing into premature completions.

\begin{figure}[h]
	\centering
	\includegraphics[width=\linewidth]{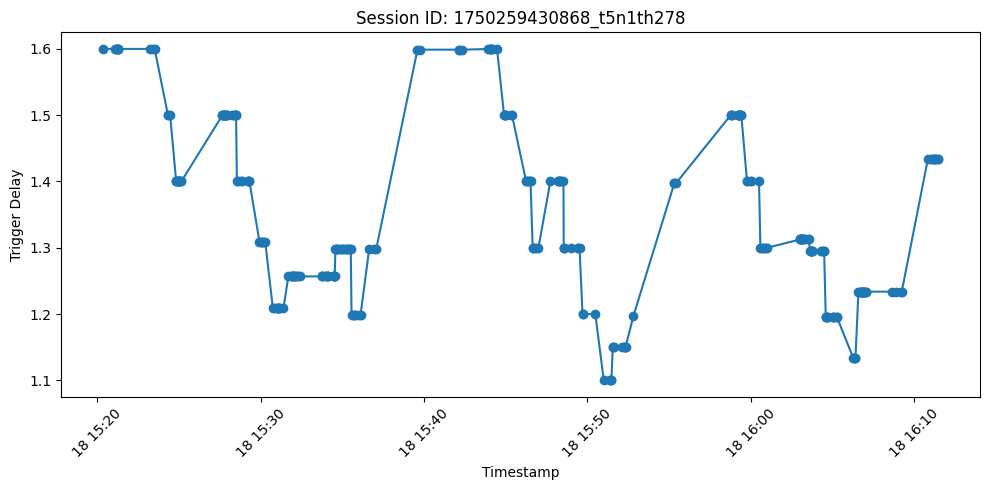}
	\caption{Effect of the acceptance rate $A$ on the computed delay $D_{\mathrm{pred}}$,  while the state anchoring ensures suggestions stay within safe cognitive bounds.}
	\label{fig:delay_session}
\end{figure}

Figure~\ref{fig:delay_session} illustrates the per-suggestion delay for one representative session. This concrete trace shows how the adaptive logic tightens or relaxes the timing in response to live feedback, while the delay range prevents disruptive jumps. The system issued 266 suggestions, of which 74 were accepted i.e. acceptance rate for this session was 27\%. The adaptive delay smoothly increased and decreased based on real-time acceptance, staying within the bounded range while avoiding abrupt swings.

\subsection{Phase-wise Evaluation of Gains}
\label{sec:evaluation}

To better understand the incremental impact of our intervention, we analyze the acceptance and blind rejection rates across the three deployment phases. We evaluate both the statistical significance and the practical implications of each transition.

\subsubsection{\textbf{Acceptance Rate Improvements}}

We compare suggestion acceptance rates using two-sample $z$-tests for proportions with pooled standard error. The results are as follows:

\begin{itemize}
	\item \textbf{Phase 1 (No delay) $\rightarrow$ Phase 2 (Static Delay):} 
	The acceptance rate increased from $p_1 = 4.89\%$ (267 accepted out of 5\,460 suggestions) to $p_2 = 15.43\%$ (189/1\,225). This yields $z = -13.22$ with $p < 10^{-38}$, indicating a highly significant improvement. Static delays thus remove many premature completions that would otherwise be dismissed.
	
	\item \textbf{Phase 2 $\rightarrow$ Phase 3 (Adaptive Delay):}
	Acceptance further increased to $p_3 = 18.60\%$ (192/1\,032), yielding $z = -2.01$ and $p = 0.045$, a statistically significant improvement from adaptive feedback alone, even after removing most mistimed suggestions through static delays.
\end{itemize}

\subsubsection{\textbf{Blind Rejection Reductions}}

Blind rejection rates declined sharply, reflecting improved alignment between suggestion timing and natural developer pauses as follows:

\begin{itemize}
	\item The blind rejection rate among all rejections dropped from $R_{\text{blind/reject}} = 8.7\%$ (453/5\,193) in Phase 1 to $1.3\%$ (13/1\,036) in Phase 2 ($z = 8.34$, $p < 10^{-16}$).
	\item It further decreased to $0.36\%$ (3/840) in Phase 3, with $z = 2.10$ and $p = 0.036$ compared to Phase 2—again, statistically significant.
\end{itemize}

\subsubsection{\textbf{Practical Impact and Cost Efficiency}}

These gains have clear real-world implications:

\begin{itemize}
	\item \textbf{Efficiency Gains:} The number of inference calls needed per accepted suggestion decreased from $20.4$ (Phase 1) to $6.5$ (Phase 2) and finally to $5.4$ (Phase 3), reflecting a 74\% reduction in backend compute per accepted completion.
	
	\item \textbf{Cost Savings:} Assuming a conservative cost of \$0.0004 per 1k tokens, the Phase~3 policy would save approximately \$7–\$8 per 1\,000 accepted completions relative to Phase~1, while also improving the developer experience.
	
	\item \textbf{Backend Stability:} Because the adaptive controller is bounded within strict delay ranges (Table~\ref{tab:delay_ranges}), it guarantees a deterministic worst-case request rate—simplifying GPU provisioning and scaling.
\end{itemize}

\subsubsection{\textbf{Validity Checks}}

All comparisons are based on sample sizes exceeding 800, ensuring reliable use of normal approximations for $z$-tests. For the smallest sample (blind rejections: 13 vs.\ 3), a Fisher’s exact test confirms the same direction and significance.

\smallskip
Our phase-wise analysis demonstrates that while static delays provide a major uplift in suggestion quality, adaptive feedback yields additional, statistically significant improvements in both acceptance and efficiency—bringing measurable benefits to both users and infrastructure.

\subsection{Role in a Broader Adaptive Ecosystem}

While our paper isolates timing control for rigorous measurement, our deployed coding assistant includes many other adaptive components. These include:
\begin{itemize}
	\item Retrieval-augmented generation (RAG) pipelines with personalized project-specific indexing.
	\item Dynamic suggestion length prediction — to avoid overwhelming developers with suggestions that are too short or too long for the current context.
	\item Suggestion ranking and reranking conditioned on user edits.
	\item Behavior-informed request suppression to prevent suggestion generation when recent signals indicate high rejection likelihood.
	
\end{itemize}

Our results confirm that timing alone can yield substantial efficiency gains. However, timing is only one axis of adaptivity. In real-world usage, these axes interact: for instance, fine-grained timing control pairs naturally with retrieval quality and length prediction to deliver suggestions that are not only well-timed but also relevant and appropriately scoped.

\subsection{Limitations}

While our field deployment demonstrates clear benefits, several practical limitations merit discussion to contextualize the results and shape future work.

\paragraph*{Generalizability across IDEs and languages.}
In our study we invited 25 developers to utilize our system in its experimental state, however, only nine voluntary power‑users were actively using the system, primarily on Python‑heavy projects. Although the timing controller is IDE‑agnostic (it requires only keystroke telemetry), interaction patterns differ in IDEs such as IntelliJ IDEA or Visual Studio, and in statically typed or compile‑centric languages (e.g., C{\small++}, Java). Broader replication across IDEs and languages is needed before large-scale deployment.

\paragraph*{Interaction effects with other adaptive levers.}
To isolate timing, we did not incorporate ranking and suggestion‑length policies.  
In production, these levers operate jointly: an aggressively shortened snippet may tolerate tighter timing, whereas a long code block might require a longer cognitive pause. Future A/B tests will examine joint optimization in a full factorial design.

\paragraph*{Proxy metrics versus real productivity.}
We used \emph{accept} versus \emph{reject} as a lightweight, privacy‑preserving proxy. Acceptance does not guarantee downstream utility: developers may accept boilerplate to edit later or reject a correct but stylistically different snippet. Subsequent work should incorporate post‑accept edit distance, compilation success, or task‑completion time to triangulate real productivity gains.

\paragraph*{Classifier drift and state granularity.}
The binary \textit{implementing/debugging} classifier, while robust, ignores nuanced modes such as refactoring or architectural exploration. Moreover, its prompt‑based inference may drift as coding workflows or LLM baselines evolve. A lightweight continual‑learning loop or periodic human‑in‑the‑loop relabeling could preserve accuracy without heavier telemetry.

\paragraph*{Long‑term habituation.}
Our few‑month window is long enough to observe steady‑state acceptance but still short relative to year‑long enterprise adoption. Developers may acclimate to smarter timing and subsequently raise their acceptance threshold. A rolling evaluation pipeline is already being deployed to monitor possible regression.

\paragraph*{Privacy and data ownership.}
Although the system transmits only minimal, aggregated signals, organizations with strict data‑sovereignty requirements may still prefer fully on‑prem inference for both the LLM and the cognitive‑state classifier. Our bounded, deterministic call‑rate design deliberately facilitates such migration by giving predictable worst‑case GPU needs.

\smallskip
Despite these caveats, the acceptance improvement together with a reduction in compute per accepted suggestion—demonstrates that timing is a first‑order lever.  Addressing the above points is likely to strengthen, not diminish, the practical value of feedback‑driven timing.  We view the present work as a foundation for a broader, multi‑axis adaptive assistant.

\subsection{Threats to Validity}
\label{sec:threats}

\textbf{Internal Validity.}  
Because participation in our deployment was voluntary, the nine active users constitute a self-selected sample.  Developers who are more open to AI tooling may be over-represented, potentially inflating acceptance rates.  We mitigated this risk by (i) advertising the plugin to \emph{all} 25 eligible engineers in the lab, (ii) analysing \emph{every} recorded interaction for those who installed it, and (iii) running the study for two full months to let initial novelty effects wear off.

\textbf{External Validity.}  
Most logged sessions were Python-centric inside VS Code.  Results may differ for languages with heavier compile steps (e.g., C\texttt{++}) or other IDEs such as IntelliJ.  However, the timing algorithm itself is IDE-agnostic and operates purely on pause durations, so porting should require only new telemetry adapters.

\textbf{Construct Validity.}  
We use \emph{acceptance rate} and \emph{blind-rejection ratio} as proxies for utility.  Acceptance is not a perfect measure of actual productivity gain—developers can accept low-quality suggestions or reject useful ones for stylistic reasons.  Prior work nonetheless shows moderate correlation between acceptance and perceived value in code-assistant studies \cite{sida2023copilot,barke2022grounded}.  We therefore focus on relative improvements across phases, which are less sensitive to any single metric’s imperfections.

\textbf{Reliability Validity.}  
Telemetry events (keystrokes, command usage) could be dropped if the IDE loses focus, and the minute-level cognitive-state classifier may drift as workflows evolve.  Both issues affect timing \emph{deterministically}: missing data lengthens delays, making our efficiency gains \emph{lower bounds}.  Classifier drift is limited by our stateless prompt-based approach; nevertheless, we plan a rolling evaluation pipeline for future deployments.

\section{Conclusion}

This work demonstrates that \emph{when} an AI code assistant chooses to surface a suggestion is as critical as \emph{what} it suggests.  By combining a logistic acceptance transform with a coarse, state signal, we devise an analytically bounded controller that is simple enough for real‑time use yet powerful enough to reshape developer behavior in the field.  In a two‑month deployment with nine professional developers, our adaptive timing raised suggestion acceptance from \textbf{4.9\%} (keystroke‑triggered baseline) to \textbf{18.6\%}, while slashing blind, instantly‑dismissed completions from \textbf{8.3\%} to \textbf{0.36\%}.  These gains cut the average number of GPU inference calls per accepted suggestion by roughly \textbf{75\%}, translating directly into lower cloud costs and energy consumption.

Beyond these concrete efficiency benefits, our results highlight three broader lessons.  
\begin{enumerate}
	\item \textbf{Timing as a design axis.} A single adaptive lever, if grounded in behavioral feedback, can yield out sized returns compared with content‑focused optimization.  
	\item \textbf{Simple state models often suffice.}  A binary \emph{implementing} vs.\ \emph{debugging} signal—derived from sparse telemetry and a compact LLM prompt—proved both robust and unobtrusive, suggesting that rich user modeling is not a prerequisite for impactful adaptivity.
	\item \textbf{Bounded policies de‑risk deployment.}  Analytical delay ranges provide deterministic worst‑case load, easing integration into production stacks where GPU capacity—and developer patience—are finite.
\end{enumerate}

\paragraph*{Future Work}  
We plan to generalize the controller across IDEs and statically‑typed languages, integrate timing with adaptive snippet‑length and retrieval policies, and replace acceptance proxies with deeper outcome metrics such as post‑accept edit distance or task‑completion time.  Longer‑horizon studies will also monitor habituation effects and classifier drift.  We envision a multi-axis, feedback-driven assistant that jointly optimizes \emph{what}, \emph{how much}, and \emph{when} to suggest—delivering code completions that are not merely accurate, but contextually and cognitively timely.

\paragraph*{Reproducibility}
An anonymized dataset and companion Jupyter notebook reproducing all quantitative analyses are available at \url{https://github.com/mohammad-nour-alawad/Optimizing-LLM-Code-Suggestions}.

\section*{Acknowledgment}
This work supported by the Ministry of Economic Development of the Russian Federation (IGK 000000C313925P4C0002), agreement No139-15-2025-010

\bibliographystyle{IEEEtran}
\bibliography{references}

\begin{thebibliography}{10}
\providecommand{\url}[1]{#1}
\csname url@samestyle\endcsname
\providecommand{\newblock}{\relax}
\providecommand{\bibinfo}[2]{#2}
\providecommand{\BIBentrySTDinterwordspacing}{\spaceskip=0pt\relax}
\providecommand{\BIBentryALTinterwordstretchfactor}{4}
\providecommand{\BIBentryALTinterwordspacing}{\spaceskip=\fontdimen2\font plus
\BIBentryALTinterwordstretchfactor\fontdimen3\font minus
  \fontdimen4\font\relax}
\providecommand{\BIBforeignlanguage}[2]{{%
\expandafter\ifx\csname l@#1\endcsname\relax
\typeout{** WARNING: IEEEtran.bst: No hyphenation pattern has been}%
\typeout{** loaded for the language `#1'. Using the pattern for}%
\typeout{** the default language instead.}%
\else
\language=\csname l@#1\endcsname
\fi
#2}}
\providecommand{\BIBdecl}{\relax}
\BIBdecl

\bibitem{brown2020language}
T.~Brown, B.~Mann, N.~Ryder, and M.~Subbiah, ``Language models are few-shot
  learners,'' in \emph{Proc. NeurIPS 2020}, vol.~33, 2020, pp. 1877--1901.

\bibitem{chen2021evaluating}
\BIBentryALTinterwordspacing
M.~Chen, J.~Tworek, H.~Jun, Q.~Yuan \emph{et~al.}, ``Evaluating large language
  models trained on code,'' 2021. [Online]. Available:
  \url{https://arxiv.org/abs/2107.03374}
\BIBentrySTDinterwordspacing

\bibitem{feng2020codebert}
Z.~Feng, D.~Guo, D.~Tang, N.~Duan, X.~Feng, M.~Gong \emph{et~al.}, ``Codebert:
  A pre-trained model for programming and natural languages.''

\bibitem{wang2021codet5}
Y.~Wang, W.~Wang, S.~Joty, and S.~C. Hoi, ``Codet5: Identifier-aware unified
  pre-trained encoder-decoder models for code understanding and generation,''
  in \emph{Proc. EMNLP 2021}, 2021.

\bibitem{wang2024systematic}
C.~Wang, S.~Gao, C.~Gao, C.~Y. Chong, S.~Gao, and M.~R. Lyu, ``A systematic
  evaluation of large code models in api suggestion: When, which, and how,'' in
  \emph{Proc. ASE 2024}, 2024, pp. 281--293.

\bibitem{li2023starcoder}
\BIBentryALTinterwordspacing
R.~Li, L.~B. allal, Y.~Zi, N.~Muennighoff \emph{et~al.}, ``Starcoder: may the
  source be with you!'' \emph{Transactions on Machine Learning Research}, 2023.
  [Online]. Available: \url{https://openreview.net/forum?id=KoFOg41haE}
\BIBentrySTDinterwordspacing

\bibitem{wang2023codellama}
\BIBentryALTinterwordspacing
B.~Rozière, J.~Gehring, F.~Gloeckle, S.~Sootla \emph{et~al.}, ``Code llama:
  Open foundation models for code,'' 2024. [Online]. Available:
  \url{https://arxiv.org/abs/2308.12950}
\BIBentrySTDinterwordspacing

\bibitem{chen2023codechain}
\BIBentryALTinterwordspacing
H.~Le, H.~Chen, A.~Saha, A.~Gokul, D.~Sahoo, and S.~Joty, ``Codechain: Towards
  modular code generation through chain of self-revisions with representative
  sub-modules,'' in \emph{The Twelfth International Conference on Learning
  Representations}, 2024. [Online]. Available:
  \url{https://openreview.net/forum?id=vYhglxSj8j}
\BIBentrySTDinterwordspacing

\bibitem{tang2023domain}
Z.~Tang, J.~Ge, S.~Liu, T.~Zhu, T.~Xu, L.~Huang, and B.~Luo, ``Domain adaptive
  code completion via language models and decoupled domain databases,'' in
  \emph{Proc. ASE 2023}, 2023, pp. 421--433.

\bibitem{li2024advpro}
X.~Li, G.~Meng, S.~Liu, L.~Xiang, K.~Sun, K.~Chen, X.~Luo, and Y.~Liu,
  ``Attribution‑guided adversarial code prompt generation for code completion
  models,'' in \emph{Proc. ASE 2024}, 2024, pp. 1460--1471.

\bibitem{feng2024complexcodeeval}
J.~Feng, J.~Liu, C.~Gao, C.~Y. Chong, C.~Wang, S.~Gao, and X.~Xia,
  ``Complexcodeeval: A benchmark for evaluating large code models on more
  complex code,'' in \emph{Proc. ASE 2024}, 2024, pp. 1895--1906.

\bibitem{sida2023copilot}
\BIBentryALTinterwordspacing
S.~Peng, E.~Kalliamvakou, P.~Cihon, and M.~Demirer, ``The impact of ai on
  developer productivity: Evidence from github copilot,'' 2023. [Online].
  Available: \url{https://arxiv.org/abs/2302.06590}
\BIBentrySTDinterwordspacing

\bibitem{barke2022grounded}
S.~Barke, M.~B. James, and N.~Polikarpova, ``Grounded copilot: How programmers
  interact with code-generating models,'' in \emph{Proc. ACM 2023}, pp.
  85--111.

\bibitem{ziegler2022productivity}
G.~Desolda, A.~Esposito, and F.~Greco, ``Understanding user mental models in
  ai-driven code completion tools: Insights from an elicitation study,''
  \emph{arXiv preprint arXiv:2502.02194v1}, 2025.

\bibitem{peng2024reposim}
C.~Peng, Q.~Wu, J.~Liu, J.~Liu, B.~Jiang, M.~Xu, Y.~Wang, X.~Liu, and P.~Yang,
  ``Reposim: Evaluating prompt strategies for code completion via user behavior
  simulation,'' in \emph{Proc. ASE 2024 (NIER Track)}, 2024, pp. 2279--2283.

\bibitem{mozannar2023when}
H.~Mozannar, G.~Bansal, A.~Fourney, and E.~Horvitz, ``When to show a
  suggestion? integrating human feedback in ai-assisted programming,'' in
  \emph{Proc. AAAI 2024}, 2024, pp. 10\,137--10\,144.

\bibitem{pu2025assistance}
K.~Pu, A.~Lazaro, I.~Arawjo, H.~Xia, and Z.~Xiao, ``Assistance or disruption?
  exploring and evaluating the design and trade-offs of proactive ai
  programming support,'' in \emph{Proc. CHI 2025}, 2025.

\bibitem{demoor2024transformer}
A.~de~Moor, A.~van Deursen, and M.~Izadi, ``A transformer-based approach for
  smart invocation of automatic code completion,'' in \emph{Proc. AIware 2024},
  2024.

\bibitem{ouyang2022training}
O.~Long, W.~Jeffrey, J.~Xu, A.~Diogo, W.~Carroll \emph{et~al.}, ``Training
  language models to follow instructions with human feedback,'' in
  \emph{Advances in Neural Information Processing Systems}, vol.~35, 2022, pp.
  27\,730--27\,744.

\bibitem{stiennon2020learning}
N.~Stiennon, L.~Ouyang, J.~Wu, and D.~M. Ziegler, ``Learning to summarize with
  human feedback,'' in \emph{Proc. NeurIPS 2020}, 2020, pp. 3008--3021.

\bibitem{chen2023promptbreeder}
C.~Fernando, D.~Banarse, H.~Michalewski, and S.~Osindero, ``Promptbreeder:
  Self-referential self-improvement via prompt evolution,'' in \emph{Proc. ICML
  2024}, 2023, pp. 13\,481--13\,544.

\bibitem{su2024dynamic}
e.~a. Su, ``When neural code completion models size up the situation,'' in
  \emph{Proc. ICSE 2024}, 2024, pp. 1--12.

\end{thebibliography}

\end{document}